\documentclass[11pt,draftcls,onecolumn]{IEEEtran}

\usepackage{graphicx}
\usepackage{amsmath}  
\usepackage{algorithmicx}
\usepackage{algpseudocode}
\usepackage{algorithm}
\usepackage{color}
\usepackage{verbatim}
\usepackage{graphicx}
\usepackage[colorinlistoftodos]{todonotes}
\usepackage{amssymb}
\usepackage{amsthm}
\usepackage{comment}
\usepackage{bm}
\usepackage{cite}
\usepackage{bbm}

\graphicspath{{../fig/}}

\newcommand{\pij}[2]{p_{#1\to#2}}

\newcommand{\pijf}[2]{p_{#1\to#2}(f)}

\newcommand{\shannon}[2]{\log\Big(1 + \frac{#1}{n_0 +  
#2}\Big)}
\newcommand{\shannonno}[2]{\log\Big(1 + \frac{#1}{n_0 
#2}\Big)}

\newcommand{\sijf}[2]{s_{#1\to#2}(f)}
\newcommand{\gij}[2]{g_{#1\to #2}}
\newcommand{\hij}[2]{h_{#1\to #2}}
\newcommand{\maximize}{\text{maximize}}
\newcommand{\ii}[0]{i_1\land i_2}

\newcommand{\Sij}[1]{S_{i\to j}(#1)}
\newcommand{\pmax}[0]{P_{\text{max}}}
\newcommand{\diff}[0]{\mathrm{d}}

\newcommand{\Nc}[0]{N_{\land}}

\newcommand{\bp}[0]{\bm{p}}
\newcommand{\indentt}[0]{{\color{white}take}} 
\newcommand{\mtR}[0]{\tilde{\mathcal{R}}}
\newcommand{\mR}[0]{\mathcal{R}}



\newtheorem{theorem}{Theorem}

\newtheorem{definition}{Definition}

\begin{document}

\title{A Resource Allocation and Coordinated Transmission Scheme for Large Cellular Networks}

\author{Jing Li and Dongning Guo\\
Department of Electrical Engineering and Computer Science\\
Northwestern University, Evanston, IL, 60208
}
\maketitle


\begin{abstract}
With the increasing number of user equipment (UE) and data demands, denser access points (APs) are being employed.
Resource allocation problems have been extensively researched with interference treated as noise. 
It is well understood that the overall spectral efficiency can be significantly improved if multiple terminals coordinate their transmissions either coherently or non-coherently.
The focus of this paper is to study how to select pairs of APs for coordination and allocate resources accordingly.
An optimization problem is formulated to maximize the network utility function by optimizing AP paring, spectrum allocation, user association, and power management in a flexible manner. 
A scalable and efficient algorithm is proposed based on iterative scheme pursuit and fractional programming.
Numerical results demonstrate substantial gains in the network utility due to coordinated transmission in a network of $128$ APs and $384$ UE.
\end{abstract}

\section{Introduction}

With dramatically increasing traffic needs, increasingly denser access points have been deployed. 
Interference has become one of the most important barriers to system capacity.
3GPP has introduced coordinated transmission since Release 11 \cite{3GPP_release11} to mitigate the effect or even make use of interference.
It is shown that coordinated transmission provides possibility for further system gain, especially for cell edge users\cite{access2010further}\cite{lee2012SpecificationSupport}. 

However, coordinated transmission is still an immature technique, many challenges need to be addressed before it can play a significant role in  practice. An important problem to settle is the flexible formation of clusters. In LTE standards, several static cell-clustering strategies with low complexity have been proposed\cite{access2010further}. However, investigation shows the performance gain based on these clustering strategies and non-coherent coordinated transmission is still relatively small, depending largely on AP deployment and traffic load\cite{brueck2010centralized}. Beyond static clustering, self-organizing networks (SON) standard and adaptive mobile-station-aware clustering concept are also proposed\cite{irmer2011coordinated}. In these strategies, APs are clustered around UEs based on their geographical position. Interference inside the cluster is decreased, but the interference caused to APs outside the cluster are not considered. This leads to possibly local and unstable benefits.

In this paper, our strategy is limited to pairwise coordinated transmission. Only cooperation between $2$ APs is allowed. Compared with coordinated transmission among many APs, this strategy is easier to implement. Analysis shows pairwise cooperation enhances system performance\cite{nigam2014coordinated}.

Besides AP coordination, other important resource management decisions include spectrum allocation, user association and power management.
Under the conventional fixed spectrum assignment policy, a significant amount of the spectrum remains unused\cite{fcc2003docket}. To use spectrum more efficiently, opportunistic dynamic spectrum allocation methods are discussed\cite{haykin2005cognitive}\cite{grandblaise2002dynamic}. Schemes based on game theory \cite{byun2017fair,wang2010toda,ji2007cognitive,ileri2005demand} or optimization\cite{ru2017power}\cite{cao2005distributed} are also well explored. However, the maximum spectrum flexibility or utilization is hard to achieve. In most formulations, subbands are predefined, APs can not dynamically split the spectrum according to UE's traffic demand or the channel state. Also, distributed methods are proposed to allocate spectrum at the expense of optimal solution.
User association is also a challenging problem in wireless network. In practice, users are associated to the AP with maximum reference SINR. 
As for power management, proper assignments not only save power but also avoid unnecessary interference. Numerous power saving models have been proposed \cite{elbatt2004joint,kubisch2003distributed,saraydar2002efficient}. However, power management problems are often non-convex. Their solutions are either computationally expensive or far from optimal. 

For resource allocation problems, Zhou and Guo proposed an algorithm \cite{zhou2018centralized} which jointly optimize spectrum allocation, user association and power management efficiently. Based on their work, this paper introduced virtual APs and formulate problems focusing on coordinated transmission. Our work seeks for the best AP pairing arrangements and the resources allocated to them.  
To achieve larger utilization, AP pairs are formed with large flexibility. Any pair of APs can choose to cooperate with each other as long as they have some UE to serve in common. Different pairing schemes can be adopted on different subbands.

In our proposed algorithms, fractional programming proposed by Shen and Yu  \cite{shen2017fractional}\cite{shen2017fractional2} plays an important role. The original problem is hard to solve due to non-convexity and the large number of continuous/integer variables. Then, 
we proved that the optimal utility can be attained by partitioning the entire spectrum into limited number of subbands, where the transmission power and cooperation scheme of all APs are constant on each subband. 
So, the key is to find these constant schemes and allocate proper amount of spectrum to them. 
With fractional programming, we are able to find new schemes efficiently by close-form iterations. 

Another feature of our model is traffic-awareness. Many previous models emphasize only on service rate, and will not make adjustments as traffic changes.  
We define the network utility function to be the average sojourn time of all packets, where the resource allocation and coordinated transmission scheme can adapt to different traffic conditions.
Notice that although this paper focuses on the quality of service, our theorems and algorithms still hold for any concave network utility function like sum rate, log sum rate, or the max-min rate.

The main contributions of this paper include:
\begin{itemize}
\item To the best of our knowledge, this is the first formulation that incorporates spectrum allocation, user association, power management and coordinated transmission  simultaneously. 

\item Coordinated AP pairings are decided with high flexibility. 
In previous papers, APs are clustered according to their geographical position. In this work, we provide network-wide solutions. APs can cooperate with the most beneficial ones, instead of the nearest ones. The pairing policy considers not only the involving APs'  spectral efficiency gain, but also their interference to their neighbors.

\item  Based on \cite{zhou2018centralized}, we prove the optimal utility can be attained by a piecewise-constant resource allocation policy.

\item 
This optimization problem is solved fast and efficiently for relatively large networks. In seeking a new resource allocation and coordinated transmission scheme, we develop an algorithm which is close-form and guarantees to converge based on fractional programming \cite{shen2017fractional}\cite{shen2017fractional2}. It fits to relatively large networks with $128$ APs and $384$ UE.
\end{itemize}

The remainder of is this paper is organized as follows. Section \ref{sec: basic model} introduces the basic resource allocation model without coordinated transmission. Section \ref{sec: problem formulation} introduces non-coherent and coherent coordinated transmission, and includes them into problem formulation. In section \ref{sec: algorithm}, we reformulate and simplify the problem. Algorithms based on Fractional Programming are proposed. Section \ref{sec: numerial results} gives numerical results. Section \ref{sec: conclustion} concludes this paper. Proofs are provided in Appendix \ref{apd: proofs}.

\section{A Basic Model without Coordinated Transmission}
\label{sec: basic model}
\subsection{Links and Resources}
Consider a network with $n$ APs, $k$ UEs, and a central controller. The central controller allocate resources and coordinate transmission on a moderate timescale. The timescale is conceived to be in the order of seconds in today's networks. On this timescale, the controller can 
keep track of the (time-averaged) channel state information (CSI) and UE traffic demands. Let the total bandwidth be $W$ Hz. 
Although channels are generally frequency selective, it is fair to assume that the spectrum resource is homogeneous on the timescale of interest, i.e., all hertz of spectrum are interchangeable.\footnote{The proposed treatment can be extended to the case of multiple independently varying subbands. See, e.g., \cite{zhou2017licensed}.} 
Denote the link from AP $i$ to UE $j $ as $i\to j$, and its gain $\gij{i}{j}$, which is flat over frequency.

In a large network, usually a UE is only served  by some nearby APs due to path loss and interference. 
Let $N = \{1,2,\ldots,n\}$ describe the set of AP indexes, $K = \{1,2,\ldots, k\}$ describe the set of UE indexes.
We define UE $j$'s neighborhood as a subset of APs that can possibly serve it. The signal-to-noise ratio of these APs should be sufficiently high and above a threshold $\xi$.
\begin{align}
    A_j = \Big\{i\;|\;i\in N, \frac{ \gij{i}{j}\pmax}{n_0} \ge \xi \Big\}.
\end{align}
where $n_0$ is the PSD of white Gaussian noise. UE $j$ treats all APs outside $A_j$ as noise. Since in practice a UE is usually served by a limited number of APs, it is fair to assume that the size of neighborhoods is upper bounded by a constant $B$. 
Also, due to physical or regulatory reasons, there is a peak power constraint for links. Let $\pmax$ denote each link's maximum possible transmission power.
\begin{align}\label{equ: power_constraint_no_coop}
    0 \le \pijf{i}{j} \le P_{\text{max}}, \quad \forall j \in K, \; i\in A_j, f \in [0,W].
\end{align}

We consider highly flexible resource allocation, 
where an AP may serve any UE at any frequency using any power (subject to power constraint) as long as the AP is in the UE's neighborhood, and a UE may be served by any sets of APs within its neighborhood. 
Let $(\pijf{i}{j}, f\in [0,W])$ denote the baseband power spectral density (PSD) AP $i$ devotes to serving UE $j$. We say AP $i$ and UE $j$ are associated if the PSD $\pijf{i}{j}$ is strictly positive over some subset of $[0,W]$ of nonzero measure. Obviously, a UE can only be associated with APs within its neighborhood.
Define the array $\bm{p}(f)$ as $(\pijf{i}{j}, f\in [0,W], j\in K, i\in A_j)$. Evidently, $\large(\bm{p}(f), f\in [0,W]\large)$ completely characterizes power spectrum allocation as well as user association in the network.

For a vector $\bm{x}$, we use $|\bm{x}|_0$ to denote its $l_0$ norm, i.e., the number of nonzero entities in  $\bm{x}$. If $\bm{x} = x$ is a scalar, then $|x|_0$ becomes the binary indicator of whether $x$ is nonzero or zero. Throughout this paper, we assume an AP may serve at most one UE at each frequency. 
The constraint can be expressed as 
\begin{align} \label{equ: zero_norm}
    \sum_{j: i\in A_j} | \pijf{i}{j} |_0 \le 1, \quad \forall i \in N, \; f \in [0,W].
\end{align}

The total transmission power of AP $i'$ is $\sum_{j':i'\in A_{j'}}\pijf{i'}{j'}$. Thus, $\gij{i'}{j}\sum_{j':i'\in A_{j'}}\pijf{i'}{j'}$ is the interference from AP $i'$ as seen by link $i\to j$.
To calculate the spectral efficiency of link $i \to j$, we use the Shannon formula:
\begin{align}\label{equ: spectral_efficiency_no_cooperate}
\sijf{i}{j} = \shannon{ \gij{i}{j}\pijf{i}{j}}{\sum_{i'\in N\setminus\{i\}}\gij{i'}{j}\sum_{j':i'\in A_{j'}}\pijf{i'}{j'}} \quad \text{bits/second/hertz}.
\end{align} 

We assume the average lengths of a packet to be \textit{L}.  The service rate of link $i \to j$ is given by 
\begin{align}\label{equ: link_rate}
   r_{i\to j} = \int_{0}^{W}\shannon{\gij{i}{j}\pijf{i}{j} }{\sum_{i' \in N\setminus\{i\} } \gij{i'}{j} \sum_{j':i'\in A_{j'}}\pijf{i'}{j'} }\diff f, \quad 
   \text{bits/second},
\end{align}
the total service rate for UE $j$ is calculated by summing up the service rates it receives from all APs.
\begin{align}\label{equ: total_user_rate}
    r_j = \sum_{i\in A_j} r_{i\to j}, \quad \forall j \in K.
\end{align}

\subsection{Utility Function}
Any general concave function 
based on the service rate vector $\bm{r} = (r_1, r_2, ..., r_k)$ can serve as the network utility function $U(\boldsymbol{r})$. 
Popular choices of function like sum rate, log sum rate, and the max-min rate are all concave.
The technology developed in this paper are applicable to general utility functions. For concreteness, we next define a specific function form for $U(\boldsymbol{r})$.
Throughout this paper, we assume the length of packets follows exponential distribution with expectation $D$ bits.
Packets intended for UE $j$ arrive at the associated APs according to a Poisson point process whose rate is $\lambda_j$ packets/second. A packet intended for UE $j$ is transmitted by all associate APs together, and it  will wait if the transmission of UE $j$'s preceding packet is incomplete.
Thus, packets for all the UEs form $k$ independent M/M/1 queues. 
With service rate $\frac{r_j}{D}$ packets/second, UE $j$'s packets have the average sojourn time of
\begin{align}
\frac{1}{(\displaystyle\frac{r_j}{D} - \lambda_j)^+},
\end{align}
where 
\begin{align}
\frac{1}{x^+} =
\begin{cases}
  \displaystyle\frac{1}{x}, & \text{if} \; x > 0, \\
  +\infty , & \text{if} \; x\leq 0.
\end{cases}
\end{align}
The network is stable if and only if the service rate is strictly greater than the the packet arrival rate for all UEs.

We define the utility function as the minus of the average sojourn time of all packets:
\begin{align}
    U(\boldsymbol{r}) = - \frac{\lambda_j}{\sum_{j\in K} \lambda_j} 
    \sum_{j\in K} 
    \frac{1}{(r_j - \lambda_j)^+}.
\end{align}
This function is a concave function in $\boldsymbol{r}$.

\subsection{A Basic Formulation without coordinated transmission}
The resource allocation problem is formulated as that of finding the PSDs $\large(\bm{p}(f), f\in [0,W]\large)$
that maximizes the network utility. 
If no coordinated transmission is allowed, the service rates are given by \eqref{equ: link_rate} and \eqref{equ: total_user_rate}. The PSDs are subject to  constraints \eqref{equ: power_constraint_no_coop} and \eqref{equ: zero_norm}. The problem is then formulated as the following, which shall be referred to as P0:
\begin{align*} 
  \underset{\boldsymbol{p, r}}{\maximize} \;\;
  & U(\boldsymbol{r}) \tag{P0a} \\ \label{P0a}
  \text{subject~to} \; \;
  & r_j =  \sum_{ i\in A_j} \int_{0}^{W}\shannon{\gij{i}{j}\pijf{i}{j} }{\sum_{i' \in N\setminus\{i\} } \gij{i'}{j}\sum_{j':i'\in A_{j'}}\pijf{i'}{j'} }\diff f, \quad
  \forall j\in K &
   \tag{P0b}\\
  & \sum_{j:i\in A_j} | \pijf{i}{j} |_0 \le 1, \quad \forall i \in N, \; f \in [0,W] &
   \tag{P0c}\\
  &0 \le \pijf{i}{j} \le P_{\text{max}}, \quad \forall  j \in K, i\in A_j, f \in [0,W]. &
   \tag{P0e}
\end{align*}

\section{A Model with Coordinated Transmission}
\label{sec: problem formulation}
\subsection{Improved Spectral Efficiency}
We shall only consider pairwise cooperation in this paper. At every frequency, an AP may either be silent, or serve one UE alone (if it is within the UE's neighborhood), or cooperate with one other AP to jointly serve one UE (if both of the APs are in the UE's neighborhood).
We consider both non-coherent coordinated transmission and coherent coordinated transmission. 
The net effect of non-coherent cooperation is basically to pool the powers to the UE they serve.
One possible implementation is to use interference cancellation techniques: The receiver decodes data from a first AP, and  eliminates the decoded signal before decoding data from a second AP\cite{irmer2011coordinated}. The net effect of coherent cooperation is that amplitudes from two APs add up at the receiver. Techniques like phase alignment bring further gain in terms of spectral efficiency. 

For ease of description, let us first explain the cooperation effect using the example of a  simple network with two APs and one UE. The UE's neighborhood includes both APs. Since there is only one UE, we omit the UE index. We also focus on a given frequency, so we also omit $f$. For example, we simplify $\pijf{1}{1}$ to $p_1$ and $\pijf{2}{1}$ to $p_2$. When APs $1$ and $2$ serve the UE in uncoordinated manner at frequency $f$, they fully interfere with each other. The total spectral efficiency is calculated as 
\begin{align}\label{equ: two_AP_no_coop}
     \shannon{p_1 g_1}{p_2 g_2}
    + \shannon{p_2 g_2}{p_1 g_1}.
\end{align}

When APs $1$ and $2$ cooperate non-coherently, their powers add up at the receiver. The total spectral efficiency of AP $1$ and $2$ is expressed as
\begin{align}\label{equ: successive decoding}
\shannonno{p_1 g_1 + p_2 g_2}{}.
\end{align}
One way to implement this is to let the two APs send independent data for successive decoding. Indeed, \eqref{equ: successive decoding} can be rewritten as
\begin{align}\label{equ: succesive decoding2}
    \shannon{p_1 g_1}{p_2 g_2} + \shannonno{p_2 g_2}{}.
\end{align}

When AP $1$ and $2$ cooperate coherently with aligned phase,\footnote{To align the phase and implement coherent combining requires overhead unaccounted for in this work.} their amplitudes add up at the receiver. The total spectral efficiency is expressed as
\begin{align}\label{equ: coherent_cooperation_2AP}
   \log\Bigg( 1+ \frac{\big(\sqrt{p_1g_1} + \sqrt{p_2g_2}\big)^2}{n_0}\Bigg).
\end{align}
From \eqref{equ: two_AP_no_coop}--\eqref{equ: coherent_cooperation_2AP}, it is easy to see that cooperation improves the spectral efficiency, and the efficiency of coherent cooperation is higher than that of non-coherent cooperation in this small network.

\subsection{Virtual APs}
We return to the general network with $n$ APs and $k$ UE. 
A pair of APs $i_1$ and $i_2$ can cooperate to serve UE $j$ over some part(s) of the spectrum if $i_1$ and $i_2$ are both in the neighborhood of $j$. 
In this case, we can think of them as a \textit{virtual} AP, denoted as $i_1\land i_2$.  
For later convenience, we assume the cooperating pair must apply identical PSDs over the said spectrum. We can think of the two cooperating links $i_1\to j$ and $i_2\to j$ together as one virtual link, denoted as $\ii \to j$.

From now on, we use $\pijf{i}{j}$ to denote the PSD AP $i$ allocates to the link $i\to j$ with no cooperation with another AP.
When APs $i_1$ and $i_2$ cooperate to serve $j$, we use $\pijf{\ii}{j}$ to denote the PSD AP $i_1$ allocates to link $i_1\to j$, which is identical to what AP $i_2$ allocates to link $i_2 \to j$. 
At the same time, we set $\pijf{i_1}{j} = \pijf{i_2}{j} = 0$ over the frequencies at which $i_1$ and $i_2$ cooperate with each other. 
If APs $i_1$ and $i_2$ do not cooperate to serve UE $j$ at frequency $f$, we set $\pijf{\ii}{j}$ to be zero.
Note that $\pijf{i}{j}$ does not always represent the PSD of the physical link $i\to j$. Instead, it is equal to the link's PSD only if $i$ works alone to serve $j$. If AP $i$ cooperates with AP $i'$ to serve UE $j$ at $f$, $\pijf{i}{j}$ is set to zero, whereas the PSD is denoted as $\pijf{i\land i'}{j}$.

Define the set of all physical APs and virtual APs as 
\begin{align}\label{equ: define Nc}
    \Nc = N\cup \{i \land i' | 1\le i < i' \le n, \exists j \;\text{s.t.}\; i\in A_j, i'\in A_j\},
\end{align}
where we let $i < i'$ in \eqref{equ: define Nc} to avoid double counting.
Note that size of each UE's neighborhood is no more than $B$. Thus, the number of virtual APs is upper bounded by $\frac{B(B-1)}{2}k$. 

With coordinated transmission, 
we redefine the neighborhood of UE $j$ to include virtual APs
\begin{align}
    A_j = \Big\{i|i\in \Nc, \frac{\gij{i}{j}\pmax }{n_0} \ge \xi \Big\}.
\end{align}
We also redefine the power vector $\bm{p}(f)$ with the new $A_j$: $\bm{p}(f) = (\pijf{i}{j}, j\in K, i\in A_j, f\in [0,W])$. Notice that in this vector, the index $i$ represents either a physical AP or a virtual AP (formed by a pair of physical APs). Likewise, for $i\in \Nc$, the link $i\to j$ represents either a physical or a virtual one. The power vector $\bm{p}(f)$ completely describes resource allocation, user association, and coordinated transmissions. The transmission power from virtual APs also follows power constraint \eqref{equ: power_constraint_no_coop}: 
\begin{align}\label{equ: power_constraint_coop}
0 \le \pijf{i}{j} \le P_{\text{max}}, \quad \forall j\in K, i\in A_j, f \in [0,W].
\end{align}

At a given frequency, a physical AP either serves a UE alone, or is silent, or coordinates with another physical AP to serve a single UE. One AP is forbidden to cooperate with two different APs at the same frequency.
Let $i < i' < i''$ denote the respective indexes of three different physical APs. Then, $\pijf{i}{j}$ and $\pijf{i \land i'}{j}$ can not both be nonzero. 
Similarly, $\pijf{i \land i'}{j}$
and $\pijf{i \land i''}{j}$ can not both be nonzero. Define 
\begin{align}\label{equ: define_involve_AP}
    N_i = \{i\}\cup \{k\land i|k<i, k\land i\in \Nc\}\cup \{i\land k| i<k, i\land k\in \Nc\},
\end{align}
which consists of physical AP $i$ and all virtual APs involving $i$. Then, these constraints are collectively represented by
\begin{align} \label{equ: zero_norm_coop}
   \sum_{i'\in N_i}\sum_{j: i'\in A_j}|\pijf{i'}{j}|_0 \le 1.
\end{align}

To better illustrate virtual APs and their constraints, consider a small network with three physical APs $\{1,2,3\}$ and two UEs $\{1,2\}$. The neighborhood of UE $1$ is $A_1 = \{1,2\}$, the neighborhood of UE $2$ is $A_2 = \{2,3\}$. The index set of all APs, including physical ones and virtual ones, is thus $\{1, 2, 3, 1\land 2, 2\land3\}$. In particular, for $i=2$, $N_i = \{1\land 2, 2, 2\land 3\}$ by \eqref{equ: define_involve_AP}, and the constraint \eqref{equ: zero_norm_coop} is expressed as 
\begin{align}
    & |\pijf{1\land 2}{1}|_0 + |\pijf{2}{1}|_0 + |\pijf{2\land 3}{1}|_0  \\
  + & |\pijf{1\land 2}{2}|_0 + |\pijf{2}{2}|_0 + |\pijf{2\land 3}{2}|_0 \le 1, 
    \quad \forall f\in [0,W].
\end{align}

Let us define
\begin{align}
    \gij{i_1\land i_2}{j} = 
    \gij{i_1}{j} + \gij{i_2}{j}, \quad i_1\land i_2 \in \Nc.
\end{align}
We next calculate a virtual AP's spectral efficiency according to Shannon formula. Consider a link $i\to j$, where AP $i\in \Nc$ can be either physical or virtual. The interference is generated by other APs, $\Nc\setminus\{i\}$. 
The transmit power of a physical AP $i'\in \Nc\setminus\{i\}$ is $\sum_{j':i'\in A_{j'}}\pij{i'}{j'}$. The interference it causes to link $i\to j$ is then $\gij{i'}{j}\sum_{j':i'\in A_{j'}}\pij{i'}{j'}$. If APs $i_3$ and $i_4$ work together as a virtual APs $i' = i_3\land i_4$, the total interference they cause to link $i\to j$ is
\begin{align}
     \gij{i_3}{j}\sum_{j':i_3\land i_4 \in A_{j'}}\pij{i_3\land i_4}{j'} + \gij{i_4}{j}\sum_{j':i_3\land i_4 \in A_{j'}}\pij{i_3\land i_4}{j'} 
= \; & \gij{i_3\land i_4}{j}\sum_{j':i_3\land i_4 \in A_{j'}}\pij{i_3\land i_4}{j'} \\
= \; & \gij{i'}{j}\sum_{j':i' \in A_{j'}}\pij{i'}{j'}.
\end{align}
To sum up, $\sum_{i'\in \Nc\setminus\{i\}}\gij{i'}{j}\sum_{j':i' \in A_{j'}}\pijf{i'}{j'}$ describes all interference to link $i\to j$ from other active APs.

Thus, when non-coherent coordinated transmission is allowed, AP $i$'s spectral efficiency is
\begin{align}\label{equ: spectral_efficiency_noncoherent}
     \sijf{i}{j}
    = \shannon{\gij{i}{j}\pijf{i}{j}}
    {\sum_{i'\in \Nc\setminus\{i\}}\gij{i'}{j}     
    \sum_{j': i'\in A_{j'}}\pijf{i'}{j'}}.
\end{align}
Notice that \eqref{equ: spectral_efficiency_noncoherent} is similar to formula \eqref{equ: spectral_efficiency_no_cooperate}, which describe the spectral efficiency of physical APs without cooperation. Virtual APs are regarded exactly the same as physical APs in terms of interference.

Likewise, for coherent coordinated transmission, the interference for each link is the same as that in \eqref{equ: spectral_efficiency_noncoherent}. In the light of \eqref{equ: coherent_cooperation_2AP}, we have the following spectral efficiency for virtual link $i_1\land i_2 \to j$:
\begin{align}
     \sijf{i_1\land i_2}{j} = \shannon{\pijf{i_1\land i_2}{j}(\gij{i_1}{j} + \gij{i_2}{j} + 2\sqrt{\gij{i_1}{j}\cdot \gij{i_2}{j}})}
    {\sum_{i'\in \Nc\setminus\{\ii\}}\gij{i'}{j}     
    \sum_{j': i'\in A_{j'}}\pijf{i'}{j'}},
\end{align}
The amount of interference remains the same regardless of coherence. This is because coherent cooperations over a given virtual link are in general non-coherent when seen by another link. 

\subsection{Problem Formulation}
The service rate contributed by virtual link $\ii \to j$ is expressed as 
\begin{align}
    r_{\ii \to j} = \int_0^W \sijf{\ii}{j} \diff f, 
    \quad \forall i_1\land i_2\in \Nc.
\end{align}
To calculate the total service rate for UE $j$, we add up the service rates for it from all physical and virtual APs.
If UE $j$ is served in coordinated manner by AP $i$ and $i'$ (both $i$ and $i'$ are in $A_j$) at frequency $f$, the service rate it receives is $r_{i\land i'\to j}(f)$. According to our rule, in this case, $\pijf{i}{j} = \pijf{i'}{j} = 0$. Thus, $r_{i\to j}$ and $r_{i'\to j}$ are zero. On the other hand, if service rate $r_{i\to j}$ is contributed by a physical AP $i$ alone ($i\in A_j$) at frequency $f$, the virtual APs that composed by $i$ all provide zero power, thus zero service rate.
Thus, the total service rate for UE $j$ is calculated as:
\begin{align}\label{equ: total_rate_coop}
    r_j = \sum_{i\in A_j}r_{i\to j}, \quad \forall j \in K.
\end{align}

Let us define the following coefficient according to whether the cooperation is coherent or non-coherent. For $i\in N$, $\hij{i}{j} = \gij{i}{j}$; for virtual AP $i\land i'\in \Nc$, 
\begin{equation}\label{equ: define h}
\hij{i\land i'}{j} =
\begin{cases}
\gij{i\land i'}{j}, \quad \text{if non-coherent} \\
\gij{i}{j} + \gij{i'}{j} + 2\sqrt{\gij{i}{j}\cdot\gij{i'}{j}}, \quad \text{if coherent.}
\end{cases}
\end{equation}

Collecting \eqref{equ: power_constraint_coop}, \eqref{equ: zero_norm_coop}, \eqref{equ: total_rate_coop}, we formulate the resource allocation and coordinated transmission problem as P1:
\begin{align*} 
    \underset{\boldsymbol{p, r}}{\maximize} \quad & U(\boldsymbol{r}) \tag{P1a}\\ 
        \text{subject~to} \quad
    & r_j = \sum_{i\in A_j} \int_0^W 
    \shannon{\hij{i}{j}\pijf{i}{j}}{\sum_{i'\in \Nc\setminus\{i\}}\gij{i'}{j} 
    \sum_{j':i\in A_{j'}}\pijf{i'}{j'}
    }\diff f, \quad \forall j\in K\tag{P1b} \label{P1b}\\
    & \sum_{i'\in N_i}\sum_{j: i'\in A_j}|\pijf{i'}{j}|_0 \le 1,
    \quad \forall i\in N, f\in [0,W] \tag{P1c} \label{P1c}\\
    & 0 \le \pijf{i}{j} \le P_{\text{max}}, \quad
    \forall j\in K, i\in A_j, f \in [0,W].
    & \tag{P1d} \label{P1d}
\end{align*}
P1 applies to both coherent and non-coherent cooperation, where the $h$ parameters are defined accordingly in \eqref{equ: define h}.

In following parts of this paper, when we mention an ``AP'', it can be either physical or virtual, unless otherwise specified.

\section{Algorithms}
\label{sec: algorithm}
Previous research \cite{zhou2018centralized} provides a solution to P0, and effectively solves the spectrum allocation, user association and power management problems. This section provides algorithms to solve P1. Compared with P0, P1 is able to identify and make the best use of cooperation opportunities by assigning proper power to virtual APs.

\subsection{Problem Reformulation}
\begin{definition}
A power allocation $(\pijf{i}{j}, j\in K, i\in A_j, f\in [0,W])$ is said to be $m$-sparse if the interval $[0,W]$ can be partitioned into $m$ sub-intervals, such that $\pijf{i}{j}$ is constant on each sub-interval for every $j\in K$ and $i\in A_j$.
\end{definition}

Following \cite{zhuang2015traffic}, we prove the following result.
\begin{theorem} \label{thm: thm1}
There exists a $k$-sparse power allocation $\bm{p}(f)$ that attains the maximum utility of P1.
\end{theorem} 

Thus, the optimal utility of P1 can be achieved by dividing the spectrum into at most $k$ subbands. Let $L$ denote the set of indexes of these subbands, $L = \{1, 2,..., k\}$. Denote the bandwidths of these subbands as $\bm{\beta} = [\beta_1, \beta_2, ..., \beta_{k}]$.  On the $l$th subband, denote the power allocation as $\bm{p}^l = (\pij{i}{j}^l, j\in K,i\in A_j)$. We can reformulate problem P1 to the following equivalent form, referred to as P2:
\begin{align*} 
    \underset{\boldsymbol{p, r, \beta}}{\maximize} \quad & U(\boldsymbol{r}) \tag{P2a}\\
    \text{subject~to}\quad
    & r_j =  \sum_{l\in L} \beta^l
    \sum_{i \in A_j}
    \shannon{\hij{i}{j}\pij{i}{j}^l}{\sum_{i' \in \Nc\setminus\{i\}}\gij{i'}{j}\sum_{j':i'\in A_{j'}}\pij{i'}{j'}^l}, 
    \quad \forall j \in K
    \tag{P2b}\\
  & \sum_{i'\in N_i}\sum_{j: i'\in A_j}|\pij{i'}{j}^l|_0 \le 1,
    \quad  \forall i \in N, l\in L
    \tag{P2c}\\
    & 0 \le \pij{i}{j}^l \le P_{\text{max}}, 
    \quad \forall j\in K, i\in A_j, l \in L
    \tag{P2d}\\
    & \beta^l \ge 0, \quad \forall l\in L 
    \tag{P2e}\\
    & \sum_{l \in L} \beta^l = W. \tag{P2e}
\end{align*}

\subsection{Algorithm for Affine Utility Function}
When the utility function is affine, it is always optimized by letting one of the $\beta_{l}$s, which contributes the maximum utility gain, equal to $W$. This policy implies a single power management scheme is implemented over the entire spectrum. 
Since each link has a constant power over the full spectrum, we denote the flat power allocated to link $i\to j$ as $\pij{i}{j}$. Denote $\bm{p} = (\pij{i}{j}, j\in K, i\in A_j)$. 

Suppose the affine utility function is expressed as $U(\bm{r}) = c_1r_1 + \ldots + c_kr_k$,  we then reformulate P2 as an equivalent problem to P3: 
\begin{align*} 
\underset{\boldsymbol{p}}{\maximize} \quad 
    & \sum_{j\in K}c_j \sum_{i \in A_j}
    \shannon{\hij{i}{j}\pij{i}{j}}{
    \sum_{i' \in \Nc\setminus\{i\}}\gij{i'}{j}\sum_{j':i'\in A_{j'}}\pij{i'}{j'}} \tag{P3a} \label{P3a}
    \\
    \text{subject~to}\quad
    & \sum_{i'\in N_i}\sum_{j:i'\in A_j}|\pij{i'}{j}|_0 \le 1,
    \quad 
     \forall i \in N \tag{P3b} \label{P3b} \\
    & 0\le \pij{i}{j} \le P_{\text{max}}, \quad \forall j\in K, i\in A_j. \tag{P3c}
\end{align*}



Equation \eqref{P3b} is equivalent to 
\begin{align}
    \sum_{i'\in N_i}\Big(\sum_{j:i'\in A_j}|p_{i'\to j}|_0 \Big) \le 1, \quad \forall i\in \Nc. 
\end{align}
We introduce auxiliary variable $d_i, i\in \Nc$. Define $d_i = \sum_{j:i\in A_j}|\pij{i}{j}|_0$, where $d_i$ is an non-negative integer. Then, constraint \eqref{P3b} is broken into two constraints:
\begin{align}
    & \sum_{j:i\in A_j}|\pij{i}{j}|_0 = d_i, \quad \forall i \in \Nc \label{equ: define_d} \\
    &  \sum_{i'\in N_i}d_{i'} \le 1, \quad \forall i \in N. \label{equ: sum_d}
\end{align}
Constraint \eqref{equ: sum_d} indicates $d_i$ is binary. Also, we write \eqref{P3a} as
\begin{align}
& \sum_{j\in K}c_j \sum_{i \in A_j}
\shannon{\hij{i}{j}\pij{i}{j}}{
\sum_{i' \in \Nc\setminus\{i\}}\gij{i'}{j}\sum_{j':i'\in A_{j'}}\pij{i'}{j'}} \notag \\
= & \sum_{i \in \Nc}\sum_{j:i\in A_j}c_j
\shannon{\hij{i}{j}\pij{i}{j}}{
\sum_{i' \in \Nc\setminus\{i\}}\gij{i'}{j}\sum_{j':i'\in A_{j'}}\pij{i'}{j'}} \\
= & \sum_{i \in \Nc} d_i \sum_{j:i\in A_j}c_j
\shannon{\hij{i}{j}\pij{i}{j}}{
\sum_{i' \in \Nc\setminus\{i\}}\gij{i'}{j}\sum_{j':i'\in A_{j'}}\pij{i'}{j'}}. \label{equ: P3a_equivalent}
\end{align}
Equation \eqref{equ: P3a_equivalent} holds because according to \eqref{equ: define_d}, when $d_i$ is zero, $\pij{i}{j} = 0$ for all $j\in K$; when $d_i$ is one, multiplying $d_i$ does not change the equation. Collecting  \eqref{equ: define_d}, \eqref{equ: sum_d} and \eqref{equ: P3a_equivalent}, we convert P3 into the following equivalent form, referred to as P4: 
\begin{align*} 
    \underset{\boldsymbol{p, d}}{\maximize} \quad 
    & \sum_{i \in \Nc}d_{i} 
    \sum_{j:i\in A_j} c_j
    \shannon{\hij{i}{j}\pij{i}{j}}{
    \sum_{i' \in \Nc\setminus\{i\}}\gij{i'}{j}\sum_{j':i'\in A_{j'}}\pij{i'}{j'}} \tag{P4a}
    \\
    \text{subject~to} \quad
    & \sum_{j:i\in A_j} |\pij{i}{j}|_0 = d_i, 
    \quad \forall i \in \Nc \tag{P4b} \label{P4b}\\
    &  \sum_{i'\in N_i}d_{i'} \le 1,
    \quad
     \forall i \in N \tag{P4c} \label{P4c}\\
    & 0 \le \pij{i}{j} \le P_{\text{max}},
    \quad \forall  j \in K, i\in A_j. \tag{P4d}
\end{align*}

We can actually view binary variable $d_{i}, i\in \Nc$ as the indicator of whether AP $i$ is active. When AP $i$ serves UE $j$ with a non-zero power, $\pij{i}{j}$ is strictly positive, binary variable $d_i = \sum_{j:i\in A_j}
|\pij{i}{j}|_0$ is equal $1$. On the other hand, when AP $i$ is silent,  $\pij{i}{j}=0$ for all $j$, so that $d_i=0$.
Notice that for a virtual AP $i = i_1 \land i_2$, $d_i = 1$ indicates physical APs $i_1$ and $i_2$ are active in a coordinated manner. 

Constraint \eqref{P4c} implies if a physical AP is active, it either works alone or cooperate with one other AP. \eqref{P4b} implies an active AP serves at most one UE. 
Let $u_{i}$ represents the UE served by AP $i$, then $u_i$ must be chosen from the set of UE whose neighborhood includes AP $i$:
\begin{align}
u_{i}\in \{j|j\in K, i\in A_j\} \cup \{0\}.
\end{align}
We define $u_{i} = 0$ if AP $i$ is off or it is on but allocate zero power to all UEs, which is possible if the interference loss caused by this AP is larger than the service rate gain it provided. 
Also, define $\gij{i}{0}=0$ and $\hij{i}{0} =0$, since  AP $i$ provides zero service rate if it is off.
With $u_i$ indicating user association, formulation P4 is equivalent to P5:
\begin{align*} 
    \underset{\boldsymbol{p, d, u}}{\maximize} \quad 
    & \sum_{i \in \Nc}d_{i} c_{u_{i}}
    \shannon{p_i\hij{i}{u_i}}{
    \sum_{i' \in \Nc\setminus\{i\}}p_{i'}\gij{i'}{u_{i}}} \tag{P5a} \label{P5a}
    \\
    \text{subject~to}\quad
    &  \sum_{i'\in N_i}d_{i'} \le 1,
     \quad 
     \forall i \in N \tag{P5b} \\
    & 0 \le p_i \le P_{\text{max}}.
    \quad \forall  i\in \Nc. \tag{P5c}
\end{align*}
Note that $p_i$ in P5 plays the role of $p_{i\to u_i}$.
P5 satisfies the definition of fractional programming.
Following work \cite{shen2017fractional} \cite{shen2017fractional2}, P5 is equivalent to P6:
\begin{align*} 
\underset{\boldsymbol{p,d,\gamma,u}}{\maximize} \quad &
 \sum_{i\in \Nc}
 c_{u_{i}} d_{i}\log(1+\gamma_{i}) -  \sum_{i\in \Nc}
 c_{u_{i}} d_{i}\gamma_{i} 
 + \sum_{i\in \Nc}
\frac{ c_{u_{i}}d_{i}(1+\gamma_{i}) p_{i}h_{i\to u_{i}}}{n_0 + \sum_{i'\in \Nc\setminus\{i\}}p_{i'}g_{i'\to u_{i}} + p_i \hij{i}{u_i}} \tag{P6a} \label{P6a}\\
  \text{subject~to}\quad
&     \sum_{i'\in N_i}d_{i'} \le 1,
    \quad 
     \forall i \in N \tag{P6b}\\
& 0 \le p_{i} \le P_{\text{max}}, 
\quad \forall i \in \Nc. \tag{P6c}
\end{align*}
Here we introduce non-negative auxiliary variables, $(\gamma_{i}, i\in \Nc)$. To understand this equivalence, notice that \eqref{P6a} is a concave function of $\boldsymbol{\gamma}$, and the utility of \eqref{P5a} and \eqref{P6a} are equal for the optimal $\boldsymbol{\gamma}$ .
For fixed $\boldsymbol{u}$, $\boldsymbol{d}$ and $\boldsymbol{p}$, we can derive the optimal $\boldsymbol{\gamma}$ by 
\begin{align} \label{equ: opt_gamma}
        \gamma^*_{i} =  \frac{p_{i}h_{i\to u_{i}}}{n_0 + \sum_{i'\in \Nc\setminus\{i\}}p_{i'}g_{i'\to u_{i}}}.
\end{align}
Then, according to \cite{shen2017fractional} \cite{shen2017fractional2}, P6 is equivalent to P7:
\begin{align*} 
\underset{\boldsymbol{p,d,\gamma,u,y}}{\maximize} \quad & 
 \sum_{i\in \Nc}
 c_{u_{i}} d_{i}\log(1+\gamma_{i}) -  c_{u_{i}} d_{i}\gamma_{i} 
 + 2y_{i}\sqrt{ c_{u_{i}} d_{i}
(1+\gamma_{i})p_{i}h_{i\to u_{i}}} \\
& - y_{i}^2(n_0 + \sum_{i'\in \Nc\setminus\{i\}}p_{i'}g_{i'\to u_{i}} + p_i \hij{i}{u_i}) \tag{P7a} \label{P7a}\\
  \text{subject~to}\quad
&    \sum_{i'\in N_i}d_{i'} \le 1,
    \quad 
     \forall i \in N \tag{P7b}\label{P7b}\\
& 0 \le p_{i} \le P_{\text{max}}, 
\quad \forall i \in \Nc. \tag{P7c} \label{P7c}
\end{align*}
where $\boldsymbol{y}$ is an array of auxiliary variables. Notice the utility function is a quadratic function of each $y_i$. When $\boldsymbol{u, d, p,\gamma}$ are fixed, we can obtain the optimal $\boldsymbol{y}$ by 
\begin{align} \label{equ: opt_y}
        y_{i}^* = \frac{\sqrt{ c_{u_{i}} d_{i}(1+\gamma_{i}) p_{i}h_{i\to u_{i}}}}{n_0 + \sum_{i'\in \Nc\setminus\{i\}}p_{i'}g_{i'\to u_{i}} + p_i \hij{i}{u_i}}.
\end{align}
Formula \eqref{P7a} can also be viewed as a quadratic function of $\sqrt{p_{i}}$. Combined with \eqref{P7b}, when all other variables are fixed the optimal $\boldsymbol{p}$ is 
\begin{align} \label{equ: opt_p}
    p_{i}^* = \min\bigg\{P_{\text{max}},\frac{c_{u_{i}} d_{i}(1+\gamma_{i}) h_{i\to u_{i}} y_{i}^2}{(\sum_{i'\in \Nc\setminus\{i\}}y_{i'}^2 g_{i\to u_{i}} + y_i^2\hij{i}{u_i})^2} \bigg\}
\end{align}

Notice that all $u_{i}$s are decoupled from each other, so we can optimize them one by one. Since the feasible region of each $u_{i}$ is finite, ${j|j\in K, i\in A_j}\cup \{0\}$, we can iterate all possible elements and find the one that maximize the utility function. Define
\begin{align}
    t_{i \to j} = & c_{j} d_{i}\log(1+\gamma_{i}) -  c_{j} d_{i}\gamma_{i} 
    + 2y_{i}\sqrt{ c_{j} d_{i}(1+\gamma_{i}) p_{i}h_{i\to j}} 
     - y_{i}^2(n_0 + \sum_{i'\in \Nc \setminus\{i\}}p_{i'}g_{i'\to j} + p_i\hij{i}{u_i}),
\end{align}
for all $j: i\in A_j$, which is the utility gain when $u_{i} = j$. We can optimize $\boldsymbol{u}$ by 
\begin{align} \label{equ: opt_u}
u_{i} = 
 \begin{cases}  
0, \; \text{if} \max_{j: i\in A_j}\{  t_{i \to j}\} < 0 \\
\text{argmax}_{j: i\in A_j} \{ t_{i \to j}\}, \; \text{otherwise},
 \end{cases}
\end{align}
where $\text{argmax}_{j\in K} \{ t_{i \to j}\}$ represents the $j$ that maximizes $t_{i\to j}$. There can be more than one $j$s that maximize $t_{i\to j}$. In this case, choosing different $j$s provides the same amount of contribution to the utility function, and we will choose the $j$ with minimal index to break the tie.

At last, we will optimize $\bm{d}$ when all other variables are fixed by a maximum weighted matching  problem in graph theory.
Let $G$ denote a graph of $n$ vertices $N = \{1,2,\ldots,n\}$ and edges $\{(i,i')| i\land i'\in \Nc\; \text{or} \; i=i'\}$. 
Each edge represents an AP: edge $(i,i'): i\land i'\in \Nc$ represents virtual AP $i\land i'$, and self-loop $(i,i)$ represents physical AP $i$. Thus, the number of edges in $G$ is no more than $n + \frac{B(B-1)}{2}k$.

A matching in a graph is a set of edges without common vertices. To decide on a feasible coordinated transmission scheme is equivalent to selecting edges for a matching. Edge $(i_1, i_2)$ is selected if and only if physical AP $i_1$ and $i_2$ cooperate; edge $(i, i)$ is selected if and only if physical AP $i$ is active and works alone. Node $i$ is not covered if and only if physical AP $i$ is silent.  To select a matching is also equivalent to assigning $1$s or $0$s to $\bm{d}$. Let $d_{\ii} = 1$ if edge $(i_1, i_2)$ is selected, $d_{i} = 1$ if edge $(i,i)$ is selected, and the remaining $d$ variables to be $0$. 

Assigning $d_{\ii}(i_1\land i_2\in \Nc)$ to be $1$ brings a utility gain calculated by \eqref{P7a}. Define the utility gain as the weight of edge $(i_1, i_2)$. Assigning $d_{i} (i\in N)$ to $1$ also brings a utility gain, which is defined as the weight of edge $(i,i)$. Thus, assigning $\bm{d}$ to maximize utility function is equivalent to selecting edges to maximize the total wight. Define
\begin{align}
v_{i} = 
& c_{u_{i}} d_{i}\log(1+\gamma_{i}) -  c_{u_{i}} d_{i}\gamma_{i} 
 + 2y_{i}\sqrt{ c_{u_{i}}(1+\gamma_{i}) p_{i}g_{i\to u_{i}}} \notag \\
& - y_{i}^2(n_0 + \sum_{i'\in \Nc\setminus\{i\}}p_{i'}g_{i'\to u_{i}} + p_i\hij{i}{u_i}), 
\quad \forall i\in \Nc. 
\end{align}
Assign $v_{\ii}$ to the edge $(i_1, i_2)$, $v_{i}$ to $(i, i)$. Then, we can use maximum weight matching algorithms to find the max-weight matching, which gives rise to $\bm{d}$ that maximizes the utility.

So far, we find the optimal value of $\boldsymbol{\gamma, y, p, u, d}$ one by one when other variables are fixed. We update theses variables in each iteration. Utility function \eqref{P7a} is guaranteed to converge because it is monotonically non-decreasing. Collecting \eqref{equ: opt_gamma}, \eqref{equ: opt_p}, \eqref{equ: opt_y} and \eqref{equ: opt_u}, we have an algorithm for resource allocation and coordinated transmission scheme for affine utility function. This algorithm does not guarantee global maximum, however, it is close-form and fast. The updating of $\boldsymbol{\gamma, y, p}$ all have time complexity of $\mathcal{O}(|\Nc|)$, which is $\mathcal{O}(n+k)$. When updating $\boldsymbol{u}$, each link $i\to j (j\in K, i\in A_j)$ is compared once. There are at most $\frac{B(B-1)}{2}k$ links in total, so, this process has time complexity $\mathcal{O}(k)$. A well-studied, mature maximum matching takes complexity $\mathcal{O}(\sqrt{|V|}|E|)$\cite{micali1980}\cite{vazirani2012improved}, where $|V|$ is the number of nodes and $|E|$ is the number of edges. Thus, updating $\boldsymbol{d}$ in graph $G$ has complexity $\mathcal{O}((m+n)\sqrt{n})$. To sum up, Algorithm I takes complexity $\mathcal{O}((m+n)\sqrt{n})$ in each iteration.

Compared with other optimization solutions, this algorithm saves computation resources. This  algorithm also serves as an important part of Algorithm 2.

\begin{algorithm}[H] 
\label{alg: affine_function}
\textbf{Initialization}\; $\boldsymbol{\gamma, y, p, u, d}$ \\
 \textbf{Repeat}\; \\
\indentt     Update $\boldsymbol{\gamma}$ by \eqref{equ: opt_gamma}\\
\indentt     Update $\boldsymbol{y}$ by \eqref{equ: opt_y}\\
\indentt     Update $\boldsymbol{p}$ by \eqref{equ: opt_p} \\
\indentt     Update $\boldsymbol{u}$ by \eqref{equ: opt_u}\\
\indentt     Update $\boldsymbol{d}$ by maximum weight matching\\
\textbf{Until}\; convergence
\caption{Resource allocation and coordinated transmission scheme for affine utility function}
\end{algorithm}
\subsection{Algorithm for Resource Allocation and coordinated Transmission}
Recall that the actual utility function in P1 is not affine. We need to decide not only $\bm{p}^l$ but also $\bm{\beta}$ to provide a resource allocation and coordinated transmission scheme over the entire spectrum. Implementing the iterative scheme pursuit algorithm \cite{zhou20171000}\cite{li2017cloud-based}, we derive the algorithm for P2. Denote the set of all patterns as $\mathcal{P}$.
\begin{algorithm}[H]\label{alg: P2} 
 \textbf{initialization}\; Find an initial feasible solution denoted by $\bm{p}_0$ with a feasible rate vector $\bm{r}_0$. Let $t \leftarrow 0, \mathcal{P} \leftarrow \{ \bm{p}_0\}$. \\
 \textbf{Repeat}\; \\
  \indentt  Solve P7 with objective function calculated by $\langle \triangledown U(\bm{r}_t), 
  \bm{r}\rangle$ 
  using Algorithm $1$, get new pattern $\bp_{t+1}$. \\
  \indentt  If $\bp_{t+1} \notin \mathcal{P}$, Let $\mathcal{P} \leftarrow \mathcal{P}\cup \{ \bp_{t+1}\}$. Otherwise, add an arbitrary pattern into $\mathcal{P}$. \\
  \indentt  Solve problem P2 to decide $\boldsymbol{\beta}$ with fixed pattern set $\mathcal{P}$.\\
\textbf{Until}\; convergence
\caption{Resource allocation and coordinated transmission scheme algorithm over the entire spectrum}
\end{algorithm}

Algorithm 2 can start with an arbitrary spectrum allocation and coordinated transmission  scheme (e.g., full spectrum reuse). In each iteration, a ``good'' pattern is identified by solving the affine utility function optimization problem by Algorithm 1, where the affine utility function is the first order approximation of current scheme. 
Intuitively, by taking the first order approximation as affine utility function, the pattern found is the most beneficial pattern for current state. This algorithm is guaranteed to converge since as the number of patterns increases, the utility function is monotonically decreasing. This algorithm does not guarantee to find the global optimal for P2, but the advantage lies on calculation speed. Also, according to numerical results in \ref{sec: numerial results}, even non-maximum solutions suffice to provide considerable performance gain for the network.

\section{Numerical Results}
\label{sec: numerial results}
\begin{figure}[!ht]
\centering
\includegraphics[clip, trim=0.5cm 9cm 0.5cm 9cm,width=\columnwidth]{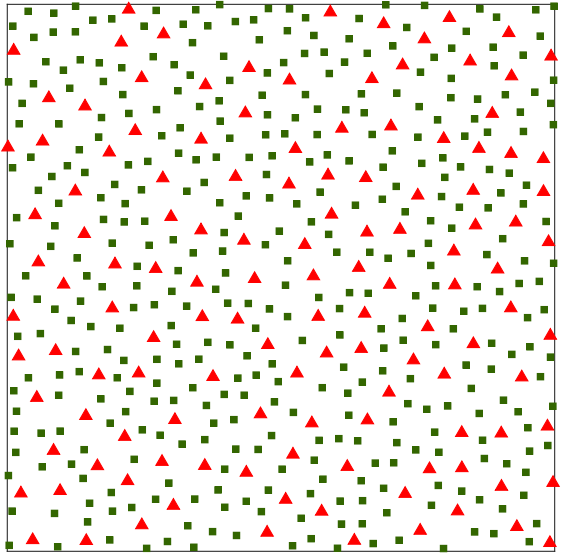}
\caption{\label{fig: topology}Topology of 128 APs and 384 UEs.}
\end{figure}

Consider a cellular network with $128$ APs and $384$ UEs with topology Fig.\ref{fig: topology}. Red triangles represent APs, green squares represent UEs. APs and UEs are randomly scattered in an area of $2400 \text{m} \times 2400 \text{m}$. The maximum transmission powers from APs are $20$ dBm. Channel gain is calculated according to LET standards \cite{access2010further}. We assume the bandwidth of the entire spectrum is $100$M hertz, and the average packet length is $1$M bits. 

\begin{figure}[!ht]
\centering
\includegraphics[width=0.7\columnwidth]{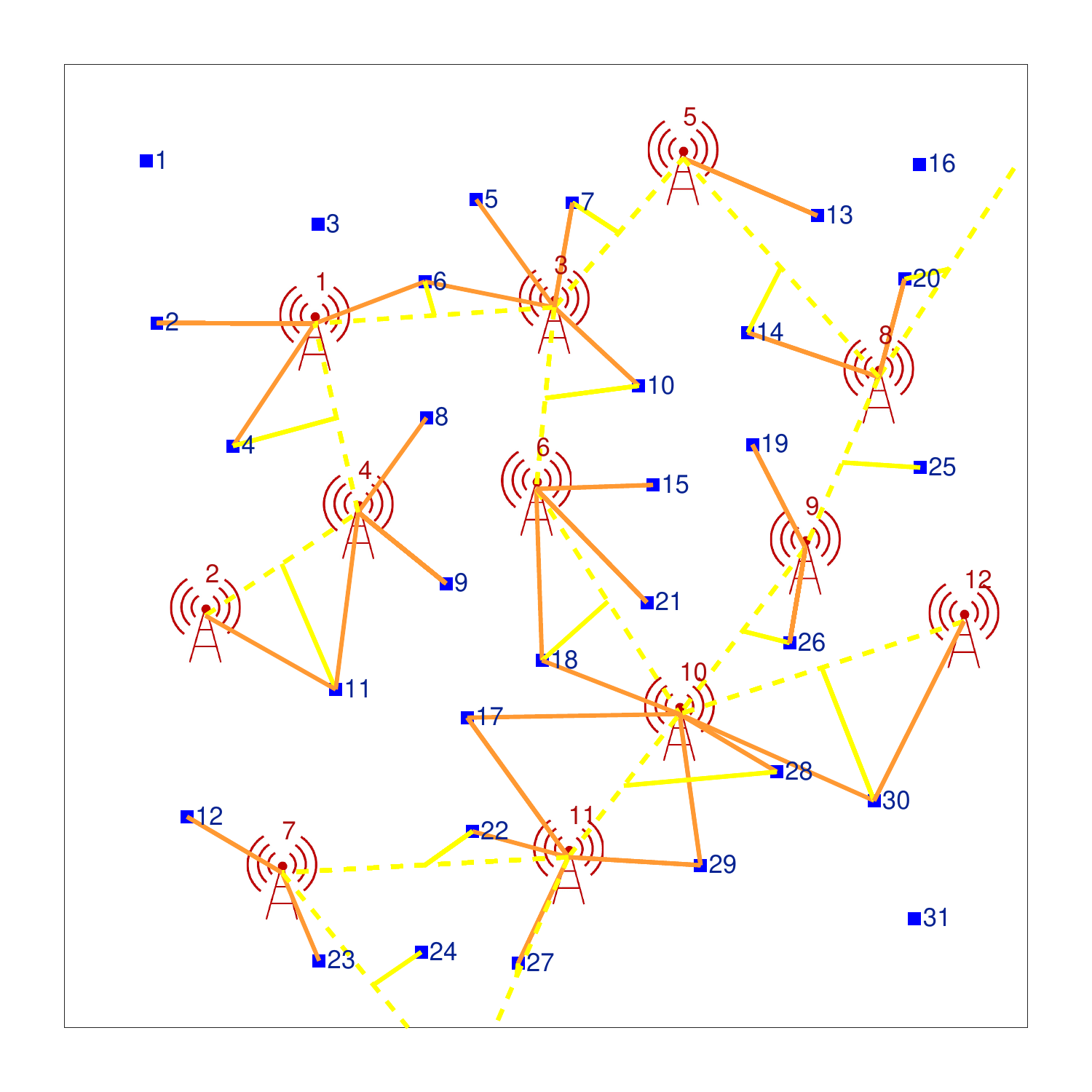}
\caption{\label{fig: topo-lines}Sectional user association and coordinated transmission details.}
\end{figure}

Fig. \ref{fig: topo-lines} illustrates user association and cooperative transmission details on a part of this topology. Red AP-like icons are APs, blue squares are UEs. An AP and a UE are associated if they are connected by an orange line. A pair of APs connected by yellow dashed line serve some UE in cooperative manner. Yellow solid lines represent coordinated transmission.

\begin{figure}[!ht]
\centering
\includegraphics[width=\columnwidth]{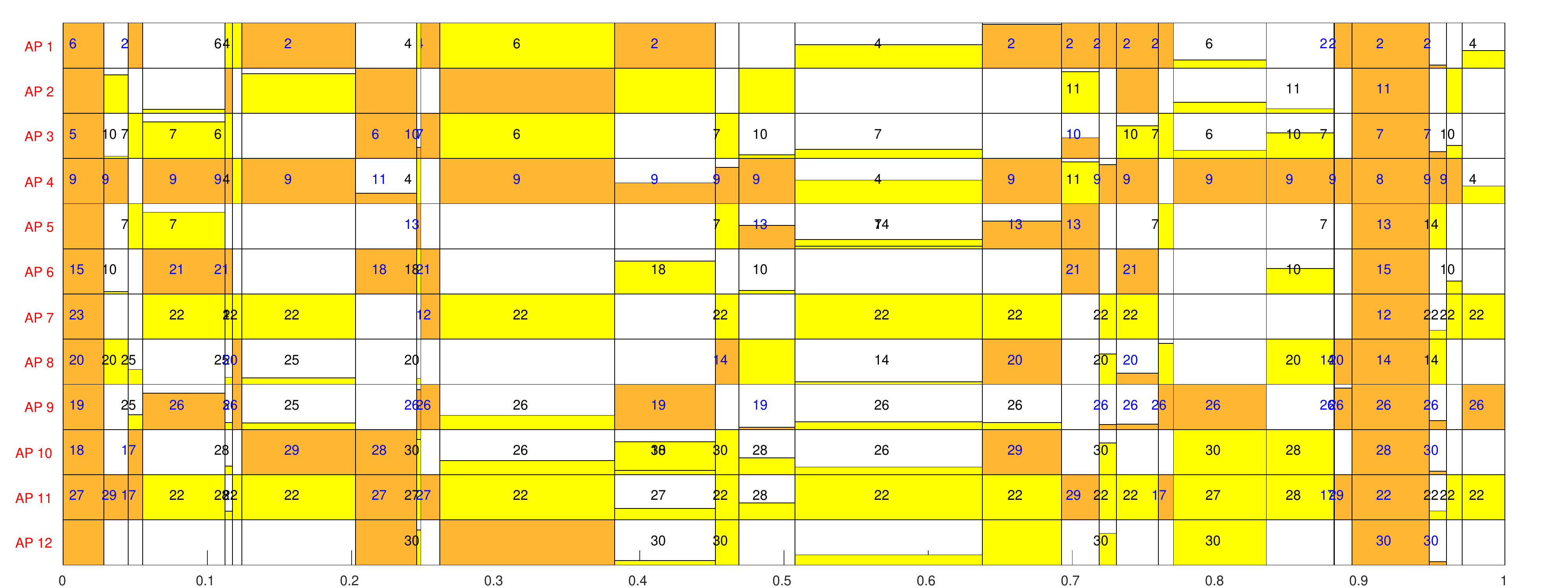}
\caption{\label{fig: block}Sectional resource allocation details.}
\end{figure}

Fig. \ref{fig: block} shows the resource allocation details of the same area of APs and UE. Each column represents a pattern, whose width is the percentage of bandwidth taken by this subband. Each row represents the behavior of an AP. Orange blocks on row $i$ and column $l$ means AP $i$ is active under subband $l$. The height of colored blocks is linear to transmission power of this link, with full height indicating full power. Text on each block indicates the index of the UE being served.
If a block is yellow, the corresponding AP serves a UE in coordinated manner. 

\begin{figure}[!ht]
\centering
\includegraphics[clip, trim=0.5cm 9cm 0.5cm 9cm,width=\columnwidth]{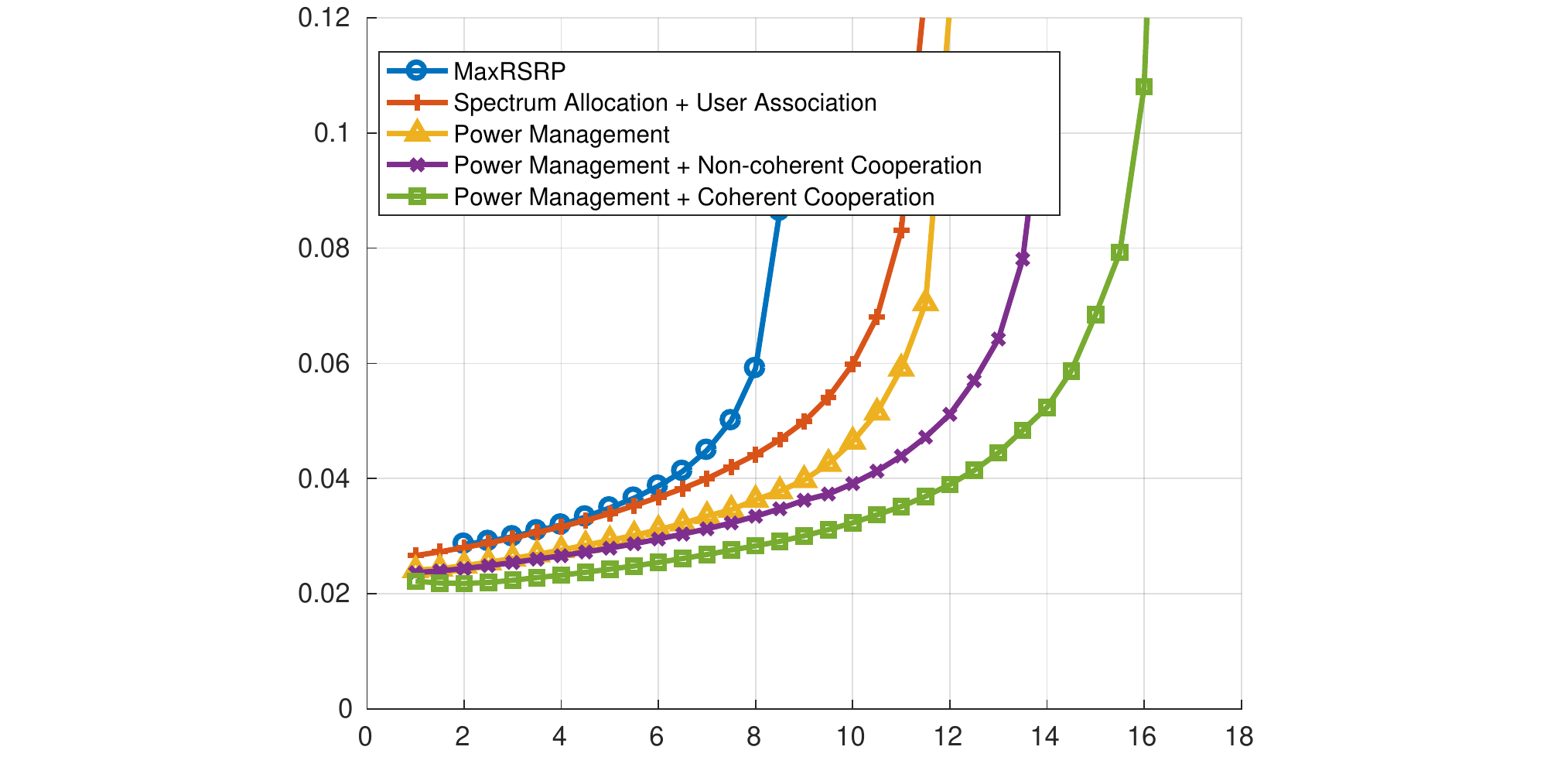}
\caption{\label{fig: performance}Topology of 128 APs and 384 UEs.}
\end{figure}

Fig.\ref{fig: performance} shows the performance improvement provided by power management and coordinated transmission. Several scenarios are compared:
\begin{enumerate}
    \item MaxRSRP: UE choose their serving APs by the maximum Reference Signal Received Power.
    \item Spectrum allocation and user association: under this scenario, the entire spectrum is split into no more than $k$ subbands. On each subband, any subset of APs are active with maximum power. Subband split and user association are optimized.
    \item power management: this scenario corresponds with formulation P0 in this paper. Using the method proposed in \cite{zhou2018centralized}, the entire spectrum is divided into no more than $k$ subbands. On each subband, any subset of APs can be active with arbitrary power. Spectrum allocation, user association, and power management are optimized.
    \item power management + non-coherent cooperation: in addition to resource allocation, pairwise non-coherent cooperation is allowed in addition to power management.
    \item power management + coherent cooperation: in addition to resource allocation, pairwise coherent cooperation is allowed in addition to power management.
\end{enumerate}

From Fig.\ref{fig: performance}, we can see the performance improves as more techniques are adopted. Under MaxRSRP user association and maximum power transmission, when the packet arrival rate is $9$ packets/second on average, the congestion time quickly reaches a cut-off point. Under spectrum allocation and power management, the arrival rate is about $11$ packets/second before the system becomes unstable. power management increased the cut-off traffic to around $12$ packets/second. Non-coherent cooperation increase the system's capacity to about $14$ packets/second, and coherent cooperation further improve the network to $16$ packets/second. In addition to spectrum allocation, user association and power management, the coordinated transmission further provides more performance gain.

\section{Conclusion}
\label{sec: conclustion}
In this paper, a comprehensive model is formulated to solve coordinated transmission, spectrum allocation, user association and power management problems. It is proved that there exists sparse optimal solutions. Our proposed algorithms are fast, efficient, and guaranteed to converge. Also, this algorithm is a successful attempt to identify the optimal pairwise clustering schemes in coordinated transmission. From numerical results, we can see flexible spectrum allocation, user association, power management and coordinated transmission give considerable gain compared with traditional MaxRSRP resource allocation.

\appendix
\section{Proof of Theorem \ref{thm: thm1}} \label{apd: proofs}
The PSDs in P1, which are Lebesgue measurable functions from an infinite dimensional function space, determine the finite dimensional rate $\textbf{r}$ and utility function. We will prove the existence of a $k$-dimensional allocation which achieves the optimal utility with Carath\'eodory's theorem.

Denote the set of Lebesgue integrable functions on $[0,W]$ as $\mathcal{L}$. Denote the number of valid links (they can be either physical or virtual) in this network as $|E|$. For ease of notation, denote the formula that calculates spectral efficiency as $\Sij{\bm{p}(f)}$:
\begin{align}
    \Sij{\bm{p}(f)} =
    \shannon{\hij{i}{j}\pijf{i}{j}}{\sum_{i'\in \Nc\setminus\{i\}}   
    \sum_{j':i'\in A_{j'}}\pijf{i'}{j'}
    \gij{i'}{j}} 
    \quad
    \forall  j\in K, i\in A_j, f\in [0,W]. 
\end{align}
Define 
\begin{align}
\begin{aligned}
R = \Big\{ & \bm{r} \in \mathbb{R}^k: \; \exists \;\textbf{p}(\cdot) = (p_{i\to j}(\cdot)) \in \mathcal{L}^{|E|}\\
& \text{s.t.}\; r_j = \sum_{i\in A_j} \int_0^W \Sij{\bm{p}(f)}\diff f,  \quad \forall j\in K,\\
    & \sum_{i'\in N_i}\sum_{j: i'\in A_j}|\pijf{i'}{j}|_0 \le 1,
    \quad \forall i\in N, f\in [0,W] \\
    & 0 \le \pijf{i}{j} \le P_{\text{max}}, \quad
    \forall j\in K, i\in A_j, f \in [0,W]
\Big\}.
\end{aligned}
\end{align}

$R$ is a subset of the $k$-dimensional Euclidean space induced by (continuous) power allocations \\ $(\pijf{i}{j})_{j\in K, i\in N_j, f\in [0,W]}$ that satisfy certain power constraints. $R$ is non-empty because it includes the zero vector $\bm{r}$ induced by all-zero power allocation.

We define another set $S$
\begin{align}
\begin{aligned}
S = \Big\{ & \bm{u} \in \mathbb{R}^k: \; \exists \;\textbf{q} \in [0,P_{\text{max}}]^{|E|}\\
& \text{s.t.}\; u_j = W \sum_{i\in A_j} \Sij{q},  \quad \forall j\in K,\\
    & \sum_{i'\in N_i}\sum_{j: i'\in A_j}|q_{i'\to j}|_0 \le 1,
    \quad \forall i\in N \Big\}.
\end{aligned}
\end{align}
The set $S$ is a subset of the $k$-dimensional Euclidean space induced by flat power allocations. It is bounded, closed, thus compact. Each vector in $S$ can be achieved when a single power profile is adopted over $[0,W]$ according to $\pijf{i}{j} = q_{i\to j}, \; \forall \; j\in K, i\in A_j, f\in [0,W]$. With this power allocation scheme, we have
\begin{align}
     r_j = \sum_{i\in A_j} \int_0^W \Sij{\bm{p}(f)}\diff f =  W \sum_{i\in A_j} \Sij{q} = u_j, \; \forall j\in K. 
\end{align}

To show that any $\bm{r}\in R$ can be attained by a $k$-sparse piecewise-constant power allocation, first, we will show that  $R$ is a subset of the convex hull of $S$, i.e., $R\subset \text{conv}(S)$. Then, we will prove that any element in $S$ can be achieved with $(k+1)$-sparse piecewise-constant power allocation.
At last, we will show that a $k$-sparse piecewise-constant power allocation suffices.

\subsection{$R$ is a subset of the convex hull of $S$}
First, the convex hull of $S$ is compact. It is easy to show the set of flat power allocations $\bm{q} \in [0,P_{\text{max}}]^{|E|}$ is compact. Since $\textbf{q} \in [0,\pmax]^{|E|}, \sum_{i'\in N_i}\sum_{j: i'\in A_j}|q_{i'\to j}|_0 \le 1,\;\forall i\in N$ is a closed subset of $[0,\pmax]^{|E|}$, it is also compact. Since $\Sij{\cdot}$ and integral are continuous functions, we know $S$ is compact. Thus convex hull of $S$ is flat. 

To show $R\subset \text{conv}(S)$, for any element $\bm{r} \in \mR$, we will construct a sequence of vectors 

$\{\bm{r}^l\}_{l = 1,2,\ldots}$ with $\bm{r}^l \in \mtR$ and $\lim_{l \to \infty} \bm{r}^l \to \bm{r}$. We will also prove $\mtR$ is compact.

$\bm{r}\in \mR$ indicates there exists a power management policy $\bm{p}(f)$ that satisfies \eqref{P1c} and \eqref{P1d} and gives service rate $\bm{r}$. Assume the PSD of each AP is a bounded measurable function of $[0,W]$.  Enumerate all links with index $0,1,\ldots, |E|-1$. 
Then, $\bm{p}(f)$ can be viewed as a function $\bm{p}: [0,W] \to [0, \pmax]^{|E|}$. $[0,W]$ is the range of frequency. Each dimension in $[0,\pmax]^{|E|}$  represents the transmission power of corresponding link.  
For a fixed $l$, we partition the $|E|$-dimensional cube $[0,\pmax]^{|E|}$ into $l^{|E|}$ disjoint cubes $\{C_{m,l}\}_{m=1}^{l^{|E|}}$, with equal length $\frac{\pmax}{l}$ at each dimension. Each cube $C_{m,l}$ can be expressed as \begin{align}
    C_m = \prod_{j\in K}\prod_{i\in A_j}[a^{m,l}_{i\to j}, a^{m,l}_{i\to j} + \frac{\pmax}{l}).
\end{align}
Define function $D(i,j)$ to map a link $i\to j$ to the index of the dimension representing its transmission power. Then, $a_{i\to j}^{m,l} = \big(\lfloor \frac{m-1}{l^{D(i,j)-1}}\rfloor \; \text{mod}\; l\big) \frac{\pmax}{l}$. We define a measurable partition of $[0,W]$ by $A_{m,l} = \bm{p}^{-1}(C_{m,l})$ and define function $\bm{p}^l: [0,W] \to [0,\max]^{|E|}$:
\begin{align}
    \pij{i}{j}^l(f) = \sum_{m=1}^{l^{|E|}}a_{i\to j}^{m,l}\mathbbm{1}_{\{ f\in A_{m,l}\}}.
\end{align}
Notice that $\bm{p}^l(f)$ is constant on any $A_{m,l}$, and $\bm{p}^l(f)$ satisfies constraints $0\le \pij{i}{j}^l(f) \le \pij{i}{j}\le \pmax$, and  $\sum_{i'\in N_i}\sum_{j: i'\in A_j}|\pij{i'}{j}^l(f)|_0 \le  \sum_{i'\in N_i}\sum_{j: i'\in A_j}|\pij{i'}{j}(f)|_0 \le 1$.

Also, we define,
\begin{align}
    r^l_j = \sum_{m=1}^{l^{|E|}} \sum_{i\in A_j}\Sij{\bm{p}^l(f)}\mu(A_{m,l}),
\end{align}
where $\mu$ is the measure function. Let $\bm{r}^l = (r^l_j, j\in K)$, it is obvious that $\bm{r}^l \in \mtR$.

From definition we know $\pij{i}{j}^l(f)$ is the closest integer multiple of $\frac{\pmax}{l}$ which is smaller or equal to $\pijf{i}{j}$.
Thus $|\pijf{i}{j} - \pij{i}{j}^l(f)| \le \frac{\pmax}{l}$, and 
\begin{align}
    |\bm{p}^l - \bm{p}| \le \sqrt{|E|}\frac{\pmax}{l}.
\end{align}

Assume $\Sij{\cdot}$ is continuous. For all $\epsilon > 0$, there exits an $\delta$, such that if $|\bm{p}_1(f) - \bm{p}_2(f)| \le \delta$, $|\Sij{\bm{p}_1(f)} - \Sij{\bm{p}_2(f)|} < \epsilon$. We let $l$ to be large enough such that $\sqrt{|E|}\frac{\pmax}{l} < \delta$. Thus, we have 
\begin{align}
    |\Sij{\bm{p}(f)} - \Sij{\bm{p}^l(f)}| < \epsilon.
\end{align}

Since $\bm{p}^l(f)$ is constant on $A_{m,l}$, we have $\sum_{m=1}^{l^{|E|}}\Sij{\bm{p}^l(f)}\mu(A_{m,l}) = \int_0^W\Sij{\bm{p}^l(f)}\diff f$. Thus,
\begin{align}
    |r_{i\to j} - r^l_{i\to j}| & 
    = |\int_0^W \Sij{\bm{p}(f)}\diff f - \sum_{m=1}^{l^{kn}}\Sij{\bm{p}^l(f)}\mu(A_{m,l})| \\
    & = \int_0^W |\Sij{\bm{p}(f)} - \Sij{\bm{p}^l(f)}| \diff f \\
    & \le \epsilon W. 
\end{align}
Thus, 
\begin{align}
    |\bm{r} - \bm{r}^l| & \le \sum_{j\in K}\sum_{i\in A_j}|r_{i\to j}-r^l_{i\to j}| \\
    & \le \epsilon n|E|W    
\end{align}
which can be arbitrarily small as $l$ is large enough. Thus, $\bm{r}^l \to \bm{r}$ as $l\to \infty$.

\subsection{$k$-sparse piecewise-constant power allocation for $\bm{r}\in \mtR$}

For flat power allocations $\bm{p}$ satisfying \eqref{P1c} and \eqref{P1d}, define the feasible region as
\begin{align}
    R_{\text{flat}} = \{ \bm{r} \in \mathbb{R}^k: r_j = \sum_{i\in A_j}W\Sij{\bm{p}}
    \}.
\end{align}

Define $\mtR$ as the convex hull of  $R_{\text{flat}}$. By carath\'eodory's theorem, any point in $\mtR$ lies in a $d$-simplex with vertices in $\mtR$.
Thus, this point can be achieved with a combination of at most $k+1$ points of $\mR_{\text{flat}}$.   

\subsection{$k$-sparse piecewise-constant power allocation for optimal solution}
Consider an optimal solution $\boldsymbol{r^*}$. According to Carath\'eodory's theorem,  $\boldsymbol{r^*}$ lies in a $d$-simplex with vertices in $\mtR$, with $d\le k$. However, since $\boldsymbol{r^*}$ is one of the optimal solutions, it can not be an interior point in this $d$-simplex (otherwise, we can get a larger utility by moving $\boldsymbol{r^*}$ with a sufficiently small distance so that the value of some dimensions are increased without harming other dimensions. This contradicts that $\boldsymbol{r^*}$ is optimal). Thus, it must lie on some $m$-face of the $d$-simplex with $m<d$, and can be written as the convex combination of $m+1 \le k$ points in $\mR_{\text{flat}}$. Therefore, an optimal solution $\boldsymbol{r^*}$ can be attained by a $k$-sparse power allocation.

\bibliographystyle{IEEEtran}
\bibliography{papers}

\begin{thebibliography}{10}
\providecommand{\url}[1]{#1}
\csname url@samestyle\endcsname
\providecommand{\newblock}{\relax}
\providecommand{\bibinfo}[2]{#2}
\providecommand{\BIBentrySTDinterwordspacing}{\spaceskip=0pt\relax}
\providecommand{\BIBentryALTinterwordstretchfactor}{4}
\providecommand{\BIBentryALTinterwordspacing}{\spaceskip=\fontdimen2\font plus
\BIBentryALTinterwordstretchfactor\fontdimen3\font minus
  \fontdimen4\font\relax}
\providecommand{\BIBforeignlanguage}[2]{{%
\expandafter\ifx\csname l@#1\endcsname\relax
\typeout{** WARNING: IEEEtran.bst: No hyphenation pattern has been}%
\typeout{** loaded for the language `#1'. Using the pattern for}%
\typeout{** the default language instead.}%
\else
\language=\csname l@#1\endcsname
\fi
#2}}
\providecommand{\BIBdecl}{\relax}
\BIBdecl

\bibitem{3GPP_release11}
T.~RAN, ``Evolved universal terrestrial radio access (e-utra) and evolved
  universal terrestrial radio access network (e-utran); coordinated multi-point
  operation for lte physical layer aspects (release 11).''

\bibitem{access2010further}
E.~U. T.~R. Access, ``Further advancements for e-utra physical layer aspects,''
  \emph{3GPP Technical Specification TR}, vol.~36, p.~V2, 2010.

\bibitem{lee2012SpecificationSupport}
J.~Lee, Y.~Kim, H.~Lee, B.~L. Ng, D.~Mazzarese, J.~Liu, W.~Xiao, and Y.~Zhou,
  ``Coordinated multipoint transmission and reception in lte-advanced
  systems,'' \emph{IEEE Communications Magazine}, vol.~50, no.~11, pp. 44--50,
  November 2012.

\bibitem{brueck2010centralized}
S.~Brueck, L.~Zhao, J.~Giese, and M.~A. Amin, ``Centralized scheduling for
  joint transmission coordinated multi-point in lte-advanced,'' in \emph{Smart
  Antennas (WSA), 2010 International ITG Workshop on}.\hskip 1em plus 0.5em
  minus 0.4em\relax IEEE, 2010, pp. 177--184.

\bibitem{irmer2011coordinated}
R.~Irmer, H.~Droste, P.~Marsch, M.~Grieger, G.~Fettweis, S.~Brueck, H.-P.
  Mayer, L.~Thiele, and V.~Jungnickel, ``Coordinated multipoint: Concepts,
  performance, and field trial results,'' \emph{IEEE Communications Magazine},
  vol.~49, no.~2, pp. 102--111, 2011.

\bibitem{nigam2014coordinated}
G.~Nigam, P.~Minero, and M.~Haenggi, ``Coordinated multipoint joint
  transmission in heterogeneous networks,'' \emph{IEEE Transactions on
  Communications}, vol.~62, no.~11, pp. 4134--4146, 2014.

\bibitem{fcc2003docket}
E.~FCC, ``Docket no 03-222 notice of proposed rule making and order,'' 2003.

\bibitem{haykin2005cognitive}
S.~Haykin, ``Cognitive radio: brain-empowered wireless communications,''
  \emph{IEEE journal on selected areas in communications}, vol.~23, no.~2, pp.
  201--220, 2005.

\bibitem{grandblaise2002dynamic}
D.~Grandblaise, D.~Bourse, K.~Moessner, and P.~Leaves, ``Dynamic spectrum
  allocation (dsa) and reconfigurability,'' in \emph{Proc. Software-Defined
  Radio (SDR) Forum}, 2002.

\bibitem{byun2017fair}
S.-S. Byun and J.-M. Gil, ``Fair dynamic spectrum allocation using modified
  game theory for resource-constrained cognitive wireless sensor networks,''
  \emph{Symmetry}, vol.~9, no.~5, p.~73, 2017.

\bibitem{wang2010toda}
S.~Wang, P.~Xu, X.~Xu, S.~Tang, X.~Li, and X.~Liu, ``Toda: Truthful online
  double auction for spectrum allocation in wireless networks,'' in \emph{New
  Frontiers in Dynamic Spectrum, 2010 IEEE Symposium on}.\hskip 1em plus 0.5em
  minus 0.4em\relax IEEE, 2010, pp. 1--10.

\bibitem{ji2007cognitive}
Z.~Ji and K.~R. Liu, ``Cognitive radios for dynamic spectrum access-dynamic
  spectrum sharing: A game theoretical overview,'' \emph{IEEE Communications
  Magazine}, vol.~45, no.~5, 2007.

\bibitem{ileri2005demand}
O.~Ileri, D.~Samardzija, and N.~B. Mandayam, ``Demand responsive pricing and
  competitive spectrum allocation via a spectrum server,'' in \emph{New
  Frontiers in Dynamic Spectrum Access Networks, 2005. DySPAN 2005. 2005 First
  IEEE International Symposium on}.\hskip 1em plus 0.5em minus 0.4em\relax
  IEEE, 2005, pp. 194--202.

\bibitem{ru2017power}
M.~Ru, S.~Yin, and Z.~Qu, ``Power and spectrum allocation in d2d networks based
  on coloring and chaos genetic algorithm,'' \emph{Procedia Computer Science},
  vol. 107, pp. 183--189, 2017.

\bibitem{cao2005distributed}
L.~Cao and H.~Zheng, ``Distributed spectrum allocation via local bargaining.''
  in \emph{SECON}, 2005, pp. 475--486.

\bibitem{elbatt2004joint}
T.~ElBatt and A.~Ephremides, ``Joint scheduling and power control for wireless
  ad hoc networks,'' \emph{IEEE Transactions on Wireless communications},
  vol.~3, no.~1, pp. 74--85, 2004.

\bibitem{kubisch2003distributed}
M.~Kubisch, H.~Karl, A.~Wolisz, L.~C. Zhong, and J.~Rabaey, ``Distributed
  algorithms for transmission power control in wireless sensor networks,'' in
  \emph{Wireless Communications and Networking, 2003. WCNC 2003. 2003 IEEE},
  vol.~1.\hskip 1em plus 0.5em minus 0.4em\relax IEEE, 2003, pp. 558--563.

\bibitem{saraydar2002efficient}
C.~U. Saraydar, N.~B. Mandayam, and D.~J. Goodman, ``Efficient power control
  via pricing in wireless data networks,'' \emph{IEEE transactions on
  Communications}, vol.~50, no.~2, pp. 291--303, 2002.

\bibitem{zhou2018centralized}
Z.~Zhou and D.~Guo, ``A centralized metropolitan-scale radio resource
  management scheme,'' \emph{arXiv preprint arXiv:1808.02582}, 2018.

\bibitem{shen2017fractional}
K.~Shen and W.~Yu, ``Fractional programming for communication systems—part i:
  Power control and beamforming,'' \emph{IEEE Transactions on Signal
  Processing}, vol.~66, no.~10, pp. 2616--2630, 2018.

\bibitem{shen2017fractional2}
------, ``Fractional programming for communication systems—part ii: Uplink
  scheduling via matching,'' \emph{IEEE Transactions on Signal Processing},
  vol.~66, no.~10, pp. 2631--2644, 2018.

\bibitem{zhou2017licensed}
Z.~Zhou, D.~Guo, and M.~L. Honig, ``Licensed and unlicensed spectrum allocation
  in heterogeneous networks,'' \emph{IEEE Transactions on Communications},
  vol.~65, no.~4, pp. 1815--1827, 2017.

\bibitem{zhuang2015traffic}
B.~Zhuang, D.~Guo, and M.~L. Honig, ``Traffic-driven spectrum allocation in
  heterogeneous networks,'' \emph{IEEE Journal on Selected Areas in
  Communications}, vol.~33, no.~10, pp. 2027--2038, 2015.

\bibitem{micali1980}
S.~Micali and V.~V. Vazirani, ``An o(v|v| c |e|) algoithm for finding maximum
  matching in general graphs,'' in \emph{21st Annual Symposium on Foundations
  of Computer Science (sfcs 1980)}, Oct 1980, pp. 17--27.

\bibitem{vazirani2012improved}
V.~V. Vazirani, ``An improved definition of blossoms and a simpler proof of the
  mv matching algorithm,'' \emph{CoRR, abs/1210.4594}, vol. 141, 2012.

\bibitem{zhou20171000}
Z.~Zhou and D.~Guo, ``1000-cell global spectrum management,'' in
  \emph{Proceedings of the 17th ACM International Symposium on Mobile Ad Hoc
  Networking and Computing}, ser. MobiHoc '17.\hskip 1em plus 0.5em minus
  0.4em\relax Chennai, India: ACM, 2017.

\bibitem{li2017cloud-based}
J.~Li and D.~Guo, ``Cloud-based resource allocation and cooperative
  transmission in large cellular networks,'' \emph{55th Annual Allerton
  Conference on Communication, Control, and Computing}, 2017.

\end{thebibliography}

\end{document}